# Sodium storage *via* single epoxy group on graphene – The role of surface doping

Nataša P. Diklić[a], Ana S. Dobrota[a,*], Igor A. Pašti[a,b], Slavko V. Mentus[a,c], Börje Johansson[b,d], Natalia V. Skorodumova[b,d]

[a]*University of Belgrade – Faculty of Physical Chemistry, Belgrade, Serbia*
[b]*Department of Materials Science and Engineering, School of Industrial Engineering and Management, KTH - Royal Institute of Technology, Stockholm, Sweden*
[c]*Serbian Academy of Sciences and Arts, Belgrade, Serbia*
[d]*Department of Physics and Astronomy, Uppsala University, Uppsala, Sweden*

**Abstract**

Due to its unique physical and chemical properties, graphene is being considered as a promising material for energy conversion and storage applications. Introduction of functional groups and dopants on/in graphene is a useful strategy for tuning its properties. In order to fully exploit its potential, atomic-level understanding of its interaction with species of importance for such applications is required. We present a DFT study of the interaction of sodium atoms with epoxy-graphene and analyze how this interaction is affected upon doping with boron and nitrogen. We demonstrate how the dopants, combined with oxygen-containing groups alter the reactivity of graphene towards Na. Dopants act as attractors of epoxy groups, enhancing the sodium adsorption on doped oxygen-functionalized graphene when compared to the case of non-doped epoxy-graphene. Furthermore, by considering thermodynamics of the Na interaction with doped epoxy-graphene it has been concluded that such materials are good candidates for Na storage applications. Therefore, we suggest that controlled oxidation of doped carbon materials could lead to the development of advanced anode materials for rechargeable Na-ion batteries.

[*]corresponding author, e-mail: ana.dobrota@ffh.bg.ac.rs

# 1. Introduction

Since its discovery in 2004, graphene has drawn wide attention as a material with novel properties related to its 2D structure. It is a single layer of $sp^2$ hybridized carbon atoms arranged in a honeycomb lattice [1]. The electronic structure of graphene is that of a zero-gap semimetal with high charge carrier mobility [2,3]. It also exhibits outstanding thermal [4], mechanical and other properties [1]. For all these reasons graphene is expected to find applications in electronics [5], energy storage [6,7], biomedicine [8] etc.

Large surface to volume ratio and high electrical conductivity of graphene make it a promising electrode material for batteries [1]. However, pristine graphene is chemically inert and interacts weakly with other chemical species through $\pi - \pi$ bonds [9], with a significant contribution of dispersion interactions in some cases [10]. Enhancing the reactivity of graphene can be achieved by disrupting its $\pi$-electron cloud, *e.g.* by introducing functional groups, dopants or defects. For instance, graphene oxide (GO), which contains at least 30 at% of oxygen [11], exhibits significantly different electronic properties and reactivity than pristine graphene. Moreover, Panchakarla *et al.* [12] demonstrated experimental procedures for synthesis of B- or N-doped graphene, and that these dopants change graphene's electronic properties. Also, dopants are known to influence the electrochemical properties of graphene [13]. After being doped with these elements, graphene turns into p-type or n-type semiconductor, respectively [12,14,15]. Therefore, the introduction of different types of defects onto the graphene basal plane can increase the number of possible applications of (functionalized) graphene. Most of the methods for producing graphene do not yield ideal, pristine, graphene (*p*-graphene), but rather graphene with structural defects [16]. For instance, if graphene is obtained through the reduction of GO, it is very hard to reduce it completely, and some of the O-groups remain at its surface, yielding reduced graphene oxide (rGO). These groups cause the corrugation of its plane due to $sp^2 \rightarrow sp^3$ C rehybridization. Thus produced rGO is hydrophilic due to the presence of oxygen-containing groups, which also enables its application in water environments, like in many electrochemical energy conversion and storage systems [6]. In this sense, metal-ion batteries are particularly interesting, as in this field graphite has been used for a long time [17], and (functionalized) graphene might be considered as a possible alternative.

Regarding the application of graphene in sodium-ion batteries, the interaction between sodium and *p*-graphene has been investigated thoroughly by the means of Density Functional Theory (DFT) calculations. It is found to be rather weak (−0.49 eV [18], −0.462 eV [19], −0.507 eV [20]) or even energetically unfavorable (0.50 eV [21]). Also, it has been proven experimentally that pristine graphene is not the best choice for the anode material in Na-ion batteries [22] and that functionalization by oxygen groups enhances Na adsorption energy [18,20]. On the other hand, substitutional doping of graphene with elements such as B or N was found to be important for tailoring the graphene reactivity towards Na [23]. Moreover, B-doped graphene is proposed as anode material in Na-ion batteries since it has been shown that it can reach capacity value which is more than 2.5 times higher than that of graphite anode in Li-ion battery [21].

While the importance of the introduction of oxygen functional groups on graphene for enhanced sodium storage is not questioned, theoretical reports suggest that different groups contribute differently to Na storage ability of a material. For example, when Na interacts with isolated OH groups on graphene basal plane NaOH formation takes place [18,20]. However, if OH groups are clustered there is no NaOH phase separation [24]. Moreover, single epoxy group acts as a stable adsorption site for Na [18,20]. Nevertheless, there is a question how much Na can be stored by a single epoxy group on graphene basal plane. Moreover, in the context of the development of novel materials for charge/metal-ion storage applications, the question is how the storage capacity of epoxy group can be modified by changing its chemical environment.

In this work, we present a DFT study on the effects of B and N substitutional doping of epoxy-graphene on its reactivity towards Na, emphasizing the possibility of its application in sodium storage systems. First, we investigate the influence of the presence of dopants on the oxygen adsorption in terms of surface functionalization. After that we investigate the interaction of doped epoxy-graphene with Na. We probe the adsorption of up to three sodium atoms in order to examine the ability of these materials to store Na.

**2. Computational details**

Periodic DFT calculations were performed within the generalized gradient approximation (GGA) using Perdew-Burke-Ernzerhof (PBE) parametrization of exchange correlation functional [25]. All the calculations were spin-polarized and carried out using PWscf code of Quantum Espresso [26], which incorporates ultra-soft pseudopotentials and plane wave basis set. The plane waves' kinetic energy cut-off was 36 Ry while the charge density cut-off was 576 Ry. Pristine graphene was modelled as one layer consisted out of 24 carbon atoms arranged in a honeycomb lattice ($C_{24}$), corresponding to the $(2\sqrt{3}\times2\sqrt{3})R30°$ structure, as done in our previous reports [23,27]. Doped graphene was modelled by substituting one C atom with B or N, yielding the dopant concentration of about 4.17 at%. A detailed description of the doped graphene models is given in one of our previous reports [27]. The k-point sampling for relaxation calculations was 4×4×1, generated using the scheme of Monkhorst and Pack [28]. For the electronic structure investigation, a denser 22×22×1 k-point grid was used. The atomic positions and cell parameters were fully relaxed until the remaining forces on atoms were below 0.002 eV Å$^{-1}$. The calculations were performed with, as well as without the correction for the long range dispersion interactions. The dispersion corrections were included using the DFT-D2 scheme of Grimme [29]. In the following text, the energies obtained using dispersion corrected calculations will be given in parentheses.

The interaction between the oxygen atom and the investigated graphene surfaces was quantified in terms of binding energies ($E_b$) calculated as:

$$E_b = E_S - (E_{X\text{-graphene}} + E_O) \tag{1}$$

where $E_S$, $E_{X\text{-graphene}}$ and $E_O$ stand for the total energy of O-functionalized graphene model surface ($p$-graphene, B-graphene or N-graphene), the total energy of graphene model surface and the total energy of isolated O atom, respectively. In the case of the interaction of $n$ Na atoms with the O-functionalized graphene surfaces we define differential adsorption energy ($E_{ads,diff}$) as:

$$E_{ads,diff} = E_{S+n\text{Na}} - (E_{S+(n-1)\text{Na}} + E_{\text{Na}}) \tag{2}$$

where $E_{S+n\text{Na}}$, $E_{S+(n-1)\text{Na}}$ and $E_{\text{Na}}$ stand for the total energy of the substrate with $n$ Na atoms, the total energy of the substrate with ($n$−1) Na atoms, the total energy of the isolated Na atom. Also, we describe the sodium interaction with functionalized graphene surfaces using the integral adsorption energies ($E_{ads,int}$) calculated as:

$$E_{ads,int} = (E_{S+n\text{Na}} - (E_S + nE_{\text{Na}}))/n \tag{3}$$

where all the quantities have the same meaning as before. The number of Na atoms interacting with functionalized graphene surfaces was increased up to 3. Clearly, the differential and integral adsorption energies are equal when $n = 1$.

The computational method for partitioning the charge density grid based on Bader algorithm [30,31] was used for charge transfer analysis. Moreover, charge redistribution caused by oxygen or sodium interaction with the modified graphene surfaces was visualized using 3D plots of charge density difference ($\Delta\rho$), defined as:

$$\Delta\rho = \rho_{SA} - \rho_{S,frozen} - \rho_A \tag{4}$$

where $\rho_{SA}$, $\rho_{S,frozen}$ and $\rho_A$ represent the ground state charge densities of the substrate interacting with the adsorbate, the frozen substrate when the adsorbate is removed, and of the isolated adsorbate in the configuration corresponding to adsorption, respectively.

## 3. Results and Discussion

*3.1. Graphene surface functionalization by oxygen*

First, we investigate the interaction of pristine, B-doped and N-doped graphene with atomic oxygen. Details on the properties of these substrates can be found in Ref. [27], where same initial models were used. Oxygen preferentially adsorbs at the C–C bridge site of *p*-graphene, forming an epoxy group on its basal plane. The equilibrium C–O bond length is 1.47 Å and the adsorption energy is −1.92 (2.02) eV, in good agreement with previous reports [32,10]. When graphene is substitutionally doped with B ($C_{23}B$), O prefers adsorbing at the C−B bridge site (Fig. 1), with the adsorption energy of −3.22 (−3.32) eV. However, in the case of N-doping ($C_{23}N$), it is more stable on the C-top site nearest to the dopant (Fig. 1), with the adsorption energy of −2.57 (−2.68) eV. Obviously, the interaction with O is much stronger in the case of doped than pristine graphene. Moreover, it should be noted that the reactivity of a large portion of basal plane is affected by the presence of dopants, but the most reactive sites are those in the vicinity of dopants and dopants themselves, in agreement with our previous study [27]. The preference of impurity sites towards the oxidation can be explained in terms of charge redistribution upon the introduction of the impurity into graphene matrix and the subsequent disruption of the π electronic system. The electronegativity of B is lower than that of C, hence some charge is transferred from B to C. Therefore, B is partially positively charged (electrophilic) and represents a favorable site for the bonding of the electron rich oxygen atom. On the other hand, N has higher electronegativity than C, and consequently is partially negatively charged in N-graphene, while the surrounding C atoms are positively charged. This results in the attraction between the electron-rich O and the first C neighbor. Bader charge analysis confirmed the charge transfer from graphene basal plane to oxygen atoms in all the three cases. The same is clearly seen in the charge difference plots presented in Fig. 1. The charge is accumulated either on the O atom (the case of epoxy-graphene and O@N-graphene system) or in the region of O–B(C) bonds (the case of O@B-doped graphene). The electronic structure properties of these surfaces will be addressed later on in more details.

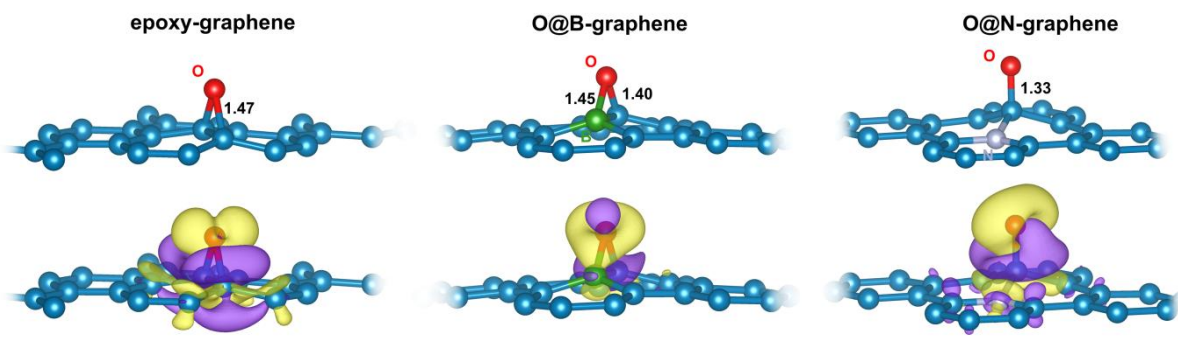

**Figure 1.** Top – the ground state structures of epoxy-graphene (left) and O-functionalized B-doped (middle) and N-doped graphene (right), with C−O and X−O (X = B or N) bond lengths given in angstroms. Bottom - the corresponding charge difference plots. Yellow isosurfaces represent charge gain, while the purple ones indicate charge loss. Isosurface values are ±0.003 e Å$^{-3}$. Graphical representations were made using VESTA [33].

Based on the above, it is clear that oxygen tends to bind close to the dopant atoms in graphene, suggesting that these sites can be considered as the sites of increased reactivity where functionalization can be performed effectively [27]. Considering our previous results on OH adsorption on doped graphene [27], the O adsorption can be considered as next oxidation step. This also reinforces previous conclusions that during the surface reduction it would be more difficult to remove oxygen functional groups from the area affected by dopants. Before proceeding to the analysis of Na bonding to O-functionalized doped graphene, we would like to emphasize the differences in the charge (re)distribution around oxygen adsorbed on B- and N-doped graphene surfaces (Fig. 1). Clearly, the chemical surroundings of an O atom on graphene basal plane tune its properties that also shows in resulting interactions with Na atoms.

*3.2. Sodium interaction with O-functionalized doped graphene surfaces*

Sodium preferentially adsorbs at the hollow site of *p*-graphene (above the center of C$_6$ hexagon), but this interaction is rather weak, with the calculated adsorption energy found between −0.35 and −0.8 eV [18-20,23,34,35], depending on the used theoretical approach. For the model of *p*-graphene used in this work, the adsorption energy of one Na atoms is −0.35 (−0.71) eV [23]. If an epoxy group is present on the graphene basal plane, Na prefers interaction with this moiety over the interaction with the basal plane. Corresponding $E_{ads}$(Na) is somewhat lower, −0.77 (−0.85) eV (Table 1), indicating a stronger interaction of Na with the oxidized graphene. Also, contribution of dispersion interaction is much higher when comparing the adsorption on *p*-graphene (0.36 eV) *versus* the interaction with epoxy-graphene (0.08 eV). Similar strengthening of Na-functionalized graphene interactions was reported by Moon et al. [20] although more exothermic Na adsorption on epoxy-graphene was predicted ($E_{ads}$(Na) = −1.02 eV). Also, we previously reported a rather strong adsorption of Na on graphene with higher concentrations of epoxy groups (~11 at.%) when single Na interacted with multiple epoxy-groups [18]. More interestingly, if an epoxy group is introduced to B- and N-doped graphene, the interaction between O and Na strengthens significantly (Table 1).

**Table 1.** Differential and integral adsorption energies (in eV) for the adsorption of 1, 2 and 3 sodium atoms on epoxy-graphene, and O-functionalized B- and N-doped graphene. The values in parentheses are obtained using PBE+D2 scheme.

|  | $C_{24}O$ (epoxy-graphene) | | $C_{23}BO$ (O@B-graphene) | | $C_{23}NO$ (O@N-graphene) | |
| --- | --- | --- | --- | --- | --- | --- |
|  | $E_{ads,diff}$ | $E_{ads,int}$ | $E_{ads,diff}$ | $E_{ads,int}$ | $E_{ads,diff}$ | $E_{ads,int}$ |
| **1Na** | −0.77 (−0.85) | | −1.53 (−1.63) | | −2.26 (−2.89) | |
| **2Na** | −2.23 (−2.28) | −1.50 (−1.87) | −2.05 (−2.72) | −1.52 (−1.89) | −1.46 (−1.97) | −1.80 (−2.09) |
| **3Na** | −0.49 (−0.63) | −1.16 (−1.46) | −1.50 (−1.91) | −1.51 (−1.90) | −1.15 (−1.26) | −1.55 (−1.85) |

While the presence of an epoxy group slightly enhances the adsorption of sodium on oxidized *p*-graphene, the dopants make it about twice (B) or three times (N) stronger, compared to pristine graphene. We also note that the interaction of Na with O-functionalized B- and N-doped graphene is stronger than the interaction of Na with the reduced forms of B- and N-doped graphene [23]. This is also in line with the observation that, for all the investigated materials, Na interacts directly with the oxygen moiety, rather than with the graphene basal plane (Fig. 2). The effect of sodium concentration is rather important for exploring how many sodium atoms can be stored by one epoxy group. Therefore, we increased the number of sodium atoms in the supercell up to three. The calculated adsorption energies are listed in Table 1, and the corresponding optimized structures are presented in Fig. 2. While on the non-doped epoxy-graphene the first sodium atom is bound rather weakly, the next one binds almost three times stronger, and the third one binds even weaker than the first one (Table 1). In the case of O@B-graphene and O@N-graphene, the strength of the interaction is dependent on the Na concentration, but it remains rather strong for up to 3 Na atoms.

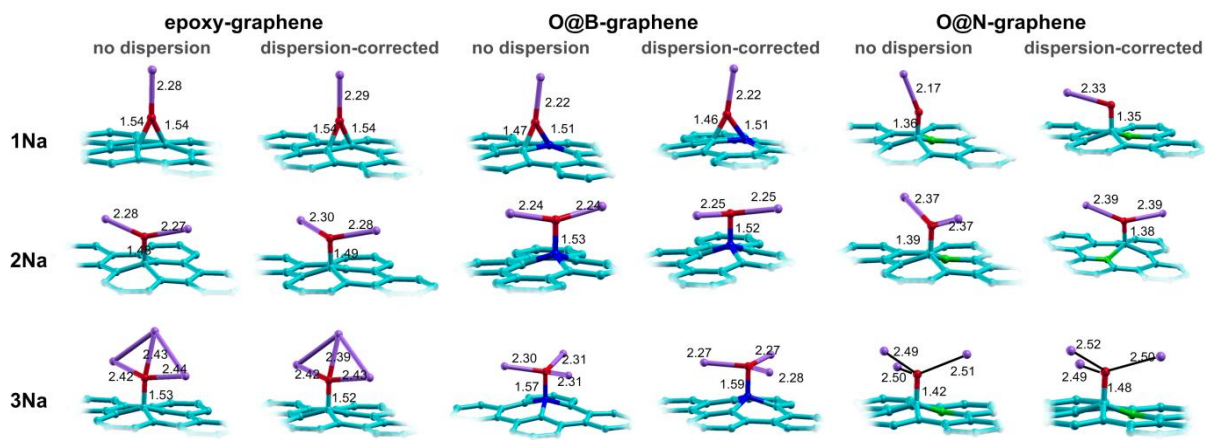

**Figure 2.** The most stable structures of 1-3 Na adsorbed on epoxy-graphene (left), O@B-graphene (middle) and O@N-graphene (right), with the bond lengths given in Ångstroms. Both PBE (no-dispersion) and PBE+D2 (dispersion-corrected) results are presented. Graphical presentations were made using XCrysden [36].

We found that Na adsorbed on epoxy-graphene transfers 0.68e to the substrate, in good agreement with Ref. [20]. For higher Na concentrations more charge is transferred from Na to the substrate. Similarly, in the case of adsorption of one Na on O@B-graphene and O@N-graphene, 0.90e and 0.99e are transferred from Na to the substrate, respectively. These results suggest that the interaction can be considered as ionic since sodium is practically in the Na$^+$ state interacting with the negatively charged epoxy group. Moreover, for higher Na concentrations charge transfer is sensitive to the number of Na atoms in the supercell. Na atoms were found to be completely ionized upon the adsorption of two sodium atoms on O@B-graphene, while when adsorbed on O@N-graphene more charge is kept by Na (0.31 to 0.36e). The situation is similar for the adsorption of the third Na atom. In order to locate where the charge goes, we have constructed 3D charge difference plots (Fig. 3). As it can be seen, the most pronounced charge rearrangement is within the oxygen-dopant complex, but also significant amount of charge is injected into the π conjugated system of the graphene basal plane around the O-dopant moiety. The charge loss in the region of O–dopant bond is an indication of its weakening upon the oxygen interaction with sodium, which is in line with the epoxide opening (cleavage of one C–O bond and switch from bridge to on top bonded O atom) and the elongation of O–dopant bond upon Na adsorption (Fig. 2). We have performed single point calculations for epoxy-graphene in the configurations corresponding to the adsorption of 1 to 3 Na and found that when 1 Na is adsorbed the deformation of the substrate (epoxy-graphene) requires 0.13 eV. However, when 2$^{nd}$ Na is adsorbed epoxide opens, which requires more than 1.01 eV. The adsorption of 3$^{rd}$ Na requires additional 0.15 eV (1.16 eV in total). The same observations hold for O@B-graphene system where epoxy-group is located between C and B atoms and opens upon the adsorption of second Na. However, in the case of O@N-graphene the maximum deformation energy is only 0.20 eV as there is no breaking of any bonds upon the adsorption of Na.

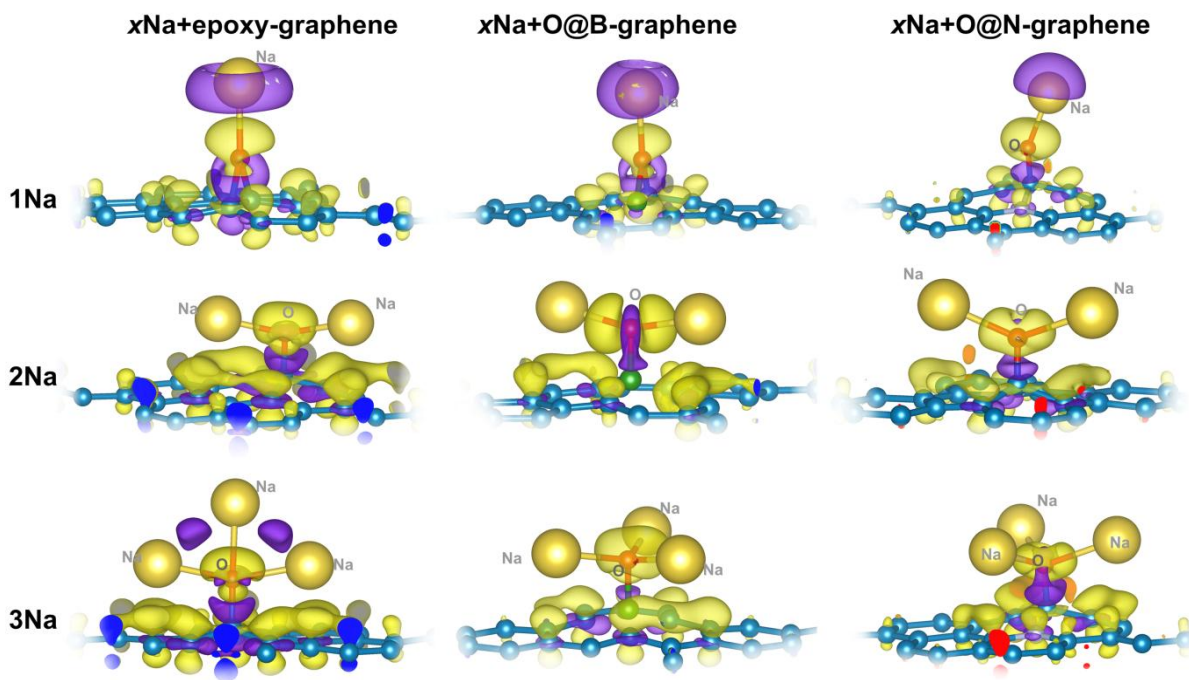

**Figure 3.** Charge difference plots for 1-3 Na adsorbed on epoxy-graphene, O@B-graphene and O@N-graphene. Yellow isosurfaces indicate charge gain, while the purple ones indicate charge loss. Isosurface values are $\pm 0.0025$ e Å$^{-3}$. Graphical representation was made using VESTA [33].

The influence of the dispersion interactions, regarding the adsorption of oxygen groups and sodium atoms, was explored by comparing the relaxed structures and adsorption energies obtained by PBE and PBE+D2 calculations. Both types of calculations yielded quite similar structures where the difference in bond length is between 0 and 0.06 Å (Fig. 2). When it comes to the adsorption energies of oxygen, a linear relation between corrected and non-corrected $E_{ads}$ was found (Fig. 4, left). In order to investigate if the observed correlation depends on the type of oxygen-containing group, we also included the data for OH adsorption on B and N-doped graphene, presented previously [27], and found that all these points fall on the same line. By linear fitting of all data points, we have found that the slope of the function is 0.99, and the intercept is −0.14 eV (and $R^2$ > 0.99). This indicates that the contribution of the dispersion interaction to the overall interaction is weakly dependent on the type of the oxygen group. Therefore, the dispersion corrected adsorption energies can be easily estimated using the values of corresponding non-corrected ones. Moreover, we previously found similar scaling in the case of Na adsorption on oxidized graphene [23] where the adsorption energies of one Na on modified graphene surfaces have been investigated. Here we confirm that there is a scaling between the PBE and PBE+D2 calculated adsorption energies (Fig. 4, right), although the scattering of the data points is much more pronounced. This is likely to be due to the fact that the interaction between Na and epoxy-graphene or O@X-graphene (X = B or N) occurs through the epoxy group, but the interactions with the basal plane can also contribute in the case of higher Na concentrations.

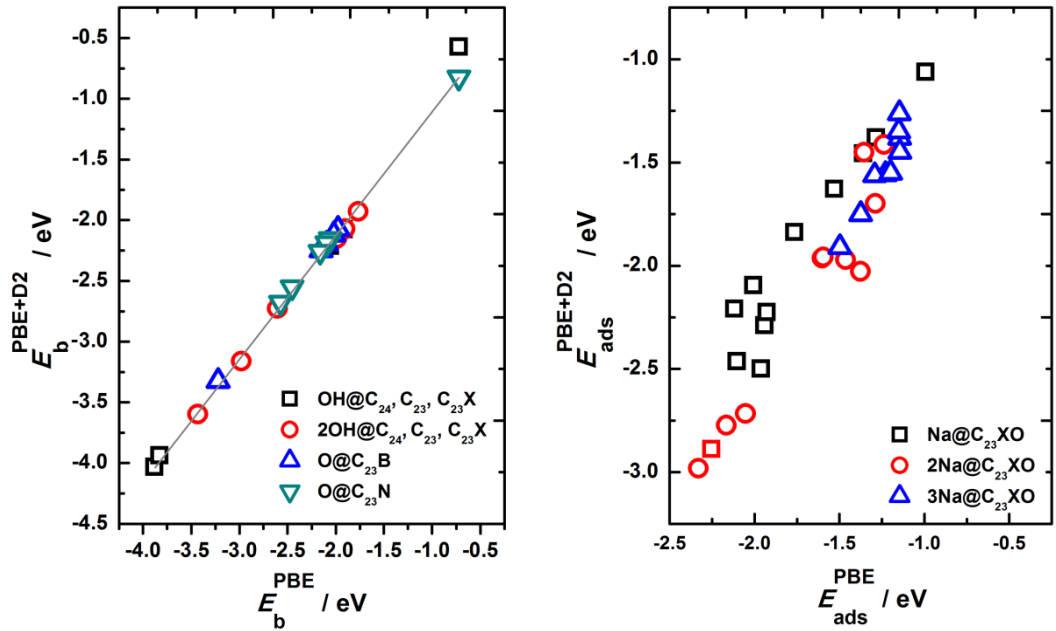

**Figure 4.** Dispersion corrected (PBE+D2) binding/adsorption energies as a function of non-corrected ones (PBE) for OH and O binding on doped graphene surfaces (left) and Na adsorption on O@X-graphene (X = B or N) (right). The data for various configurations of adsorbates on graphene surfaces are used.

*3.3. Doped and non-doped epoxy-graphene as an electrode material in sodium storage*

We consider two criteria which determine the possibility of using a given material as an electrode material in metal-ion batteries: (i) its electrical conductivity; and (ii) its stability under the working conditions (considering possible phase separations and metal precipitation). It is known from previous literature reports that heavy oxidation of graphene results in band-gap opening [37] due to the disruption of the π electronic system and the corrugation of the graphene basal plane due to $sp^2 \rightarrow sp^3$ rehybridization of C atoms bound to oxygen functional groups. Hence, in order to address possible conductivity issues of the investigated materials, we inspected their electronic densities of states (DOS, Fig. 5). At a relatively low O concentrations used in our work (< 5 at.%), no band-gap opening was observed for $C_{23}O$, $C_{23}BO$ and $C_{23}NO$ models (Fig. 5). It is of great importance that this situation holds during Na adsorption, as an electrode material should be conducting at every stage of the cell operation. Therefore, we have also inspected the DOSes of the studied systems upon Na adsorption and no band gap opening was observed for any of the investigated systems.

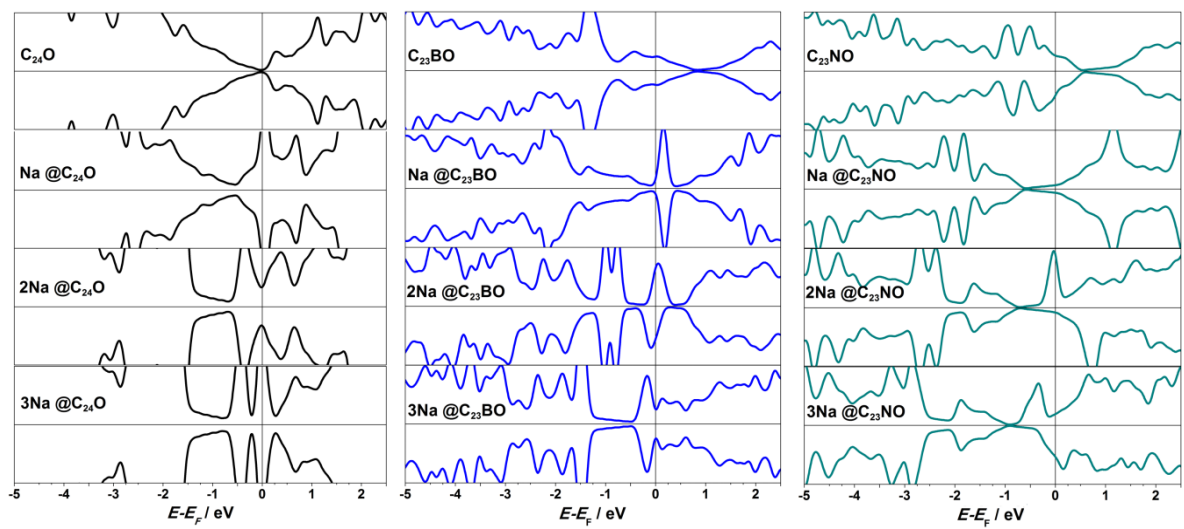

**Figure 5.** Densities of states of epoxy-graphene ($C_{24}O$), O@B-graphene ($C_{23}BO$) and O@B-graphene ($C_{23}NO$) before (top row) and upon the adsorption of 1 (2$^{nd}$ row), 2 (3$^{rd}$ row) or 3 (4$^{th}$ row) Na. The energies are reported with respect to the Fermi energy ($E_F$), which is set to 0 and indicated by vertical line.

Regarding the stability questions we next discuss possible phase separation processes upon sodiation of the studied materials. In contrast to the case of single OH groups on graphene basal plane [18,24], no formation of Na-oxygen phases was observed in any case (Fig. 2). Moreover, upon removal of Na from the optimized structures and the re-relaxation of Na-free systems the initial configurations are restored, suggesting that the studied materials should be stable upon the extended sodiation/de-sodiation cycles, like in a rechargeable Na-ion battery. This is very important for good and long battery performance, which requires reversible metal-electrode material interactions.

Another unfavorable possibility is metal precipitation, which can be discussed by comparing the obtained Na adsorption energies with the cohesive energy of metallic Na (−1.13eV [38], Fig. 6). The Na adsorption energies should be lower than its cohesive energy. Otherwise, sodium precipitation will occur, which could significantly impact the battery performance. As it can be seen from Table 1, all the calculated Na adsorption energies are below the Na cohesive energy, except for the case of one sodium atom on epoxy-graphene (Fig. 6). Even in the case of 2 Na adsorbed at single epoxy-group on graphene differential adsorption energy of 3$^{rd}$ Na is very small and the metal phase would is preferred over atomic adsorption on epoxy-group. In fact, it can be seen that in this case there is direct interaction between 3 Na atoms which might indicate the first step of metal precipitation. Therefore, we conclude that the non-doped epoxy-graphene with low concentration of epoxy-groups has limited possibilities for battery applications. However, in Ref. [18] it is found that this is not the case for higher concentrations of epoxy groups on graphene suggesting that the performance can be boosted by precisely controlled oxidation of graphene. In contrast to non-doped epoxy-graphene, the doping with B and N improves the Na storage ability and prevents Na precipitation (Fig. 6). Using the approach described in Ref. [23] we calculated sodiation potentials for the investigated system, which are reported in Fig. 6. In conclusion, we suggest that the performance of non-doped epoxy-graphene as an electrode material in sodium-ion battery will strongly depend on its degree of oxidation, while the introduction of dopants into epoxy-graphene or controlled oxidation of doped graphene can greatly enhance Na storage capabilities of these materials.

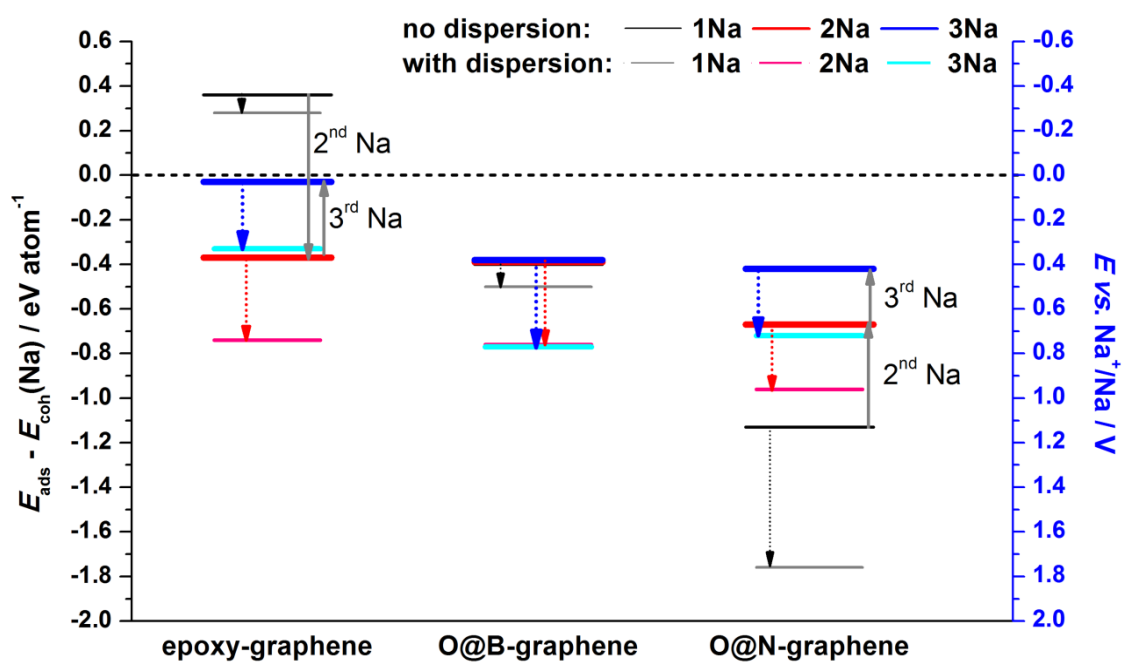

**Fig 6.** Dispersion uncorrected (PBE) and corrected (PBE+D2) Na adsorption energies on non-doped epoxy-graphene, O@B-graphene and O@N-graphene, referred with respect to the Na cohesive energy. On the right scale sodiation potentials are evaluated following Ref. [23]. Integral Na adsorption energies were used for the calculations.

**4. Conclusion**

We have studied the interaction of Na with oxygen functionalized B- and N-doped graphene, as well as with non-doped epoxy-graphene. We find that the considered dopants act as attractors of epoxy group and that the binding of oxygen is stronger on doped surfaces, compared to pristine graphene. Also, the presence of N and B contributes significantly to the reactivity of graphene towards Na. It is shown that the interaction between Na and oxidized doped graphene is stronger compared to the case of non-doped epoxy-graphene. Furthermore, these materials show the ability to store up to, at least, three Na while the analysis of their electronic structures indicates no band gap opening, which is in favor of their application as an electrode material in Na-ion batteries. Moreover, oxygen-functionalized B- and N-doped graphene are expected to provide stable operation without metal precipitation under sodiation/de-sodiation cycles. While the presented results point to possible strategies for improving metal/charge storage capabilities of graphene based materials, the presented work also emphasizes the need to consider the co-existence of dopants/impurities with the oxygen functional groups on graphene basal plane when discussing their electrochemical properties.

**Acknowledgement**

A.S.D, I.A.P. and S.V.M. acknowledge the support provided by the Serbian Ministry of Education, Science, and Technological Development (III45014). N.V.S. acknowledges the support provided by Swedish Research Council through the project No. 2014-5993. We also acknowledge the support from Carl Tryggers Foundation for Scientific Research. The computations were performed on resources provided by the Swedish National Infrastructure for Computing (SNIC) at High Performance Computing Center North (HPC2N) at Umeå University.